\documentclass[12pt]{iopart}

\usepackage{graphicx} 
\usepackage{color}
\usepackage{url}

\begin{document}

\title[]{Exploring CP-even scalars of a Two Higgs-doublet model in future $e^-\,p$ colliders}

\author{Chuene Mosomane$^{a, 1}$, Mukesh Kumar$^{a, 2}$,  Alan S. Cornell$^{a, 3}$ and Bruce Mellado$^{b, 4}$}
\address{$^{a}$ National Institute for Theoretical Physics; School of Physics and Mandelstam Institute for
Theoretical Physics, University of the Witwatersrand, Johannesburg, Wits 2050, South Africa.}
\address{$^{b}$ School of Physics, University of the Witwatersrand, Johannesburg, Wits 2050, South Africa.}
\ead{$^{1}$cjmosomane@gmail.com, $^{2}$mukesh.kumar@cern.ch, $^{3}$alan.cornell@wits.ac.za, $^{4}$bruce.mellado@wits.ac.za}
\vspace{10pt}

\begin{abstract}
In this proceeding we shall explore the potential of a future $e^- \,p$ collider to probe the 
CP-even scalars in a two Higgs doublet model. We consider Type-I in this study. The mass 
of the lighter scalar particle is considered to be the Higgs-boson, $m_h = 125$~GeV, and a 
heavy scalar $m_H = 270$~GeV. The centre of mass energy for the $e^-\,p$ collision is 
considered as in the Large Hadron Electron Collider and the Future Circular Hadron 
Electron Collider configurations, by fixing the proton beam energy to be $E_p = 7$ and 
50~TeV, respectively, and an electron beam energy of $E_e = 60$~GeV. Production cross 
sections of these scalars are also shown at higher electron beam energies. 
Future prospects of these studies are also discussed.   
\end{abstract}

\section{Introduction}
\label{intro}
After the discovery and confirmation of the Standard Model (SM) Higgs boson ($h$) at the 
Large Hadron Collider (LHC)~\cite{Aad:2012tfa, Chatrchyan:2012xdj}, it is pragmatic to 
expect hints of any beyond the SM (BSM)
particle spectrum by using the present energies in the high luminosity runs. The hints of 
BSM physics in current and coming data sets from the LHC can be seen in distortions of 
differential distributions of kinematics of the Higgs boson or any other excesses in data by 
considering different final states, as noted in 
Refs.~\cite{vonBuddenbrock:2015ema, vonBuddenbrock:2016rmr, Kumar:2016vut}. 
In these references the authors considered the two-Higgs doublet model (THDM) as a next 
step in searches of BSM heavy scalars, with a possible spectrum and parameter space at 
the LHC. In this proceeding we will focus on CP-even scalars $h$ and $H$ of the THDM, 
and study the possibilities of these scalars at future $e^-\,p$ colliders, such as the Large Hadron 
electron Collider (LHeC) and the future circular hadron electron collider (FCC-he).

In Section~\ref{2hdm} we briefly discuss the THDM, after which we provide a preliminary study 
on cross section measurements of $h$ and $H$ at the LHeC and FCC-he energies. Future 
detailed prospects for these studies are pointed out in Section~\ref{future}. 
\section{Two-Higgs doublet model}
\label{2hdm}
It is well known that in the SM $SU(2)_L \otimes U(1)_Y$ electroweak theory that the Higgs sector consists
of only one complex Higgs doublet, and hence only one physical neutral Higgs scalar boson $h$ (where
its mass is a free parameter and not fixed by theory). Recall that Run-I and Run-II data at the LHC confirms 
its mass, $m_h \approx 125$~GeV~\cite{Aad:2012tfa, Chatrchyan:2012xdj}. If we consider two complex 
$SU(2)_L$ doublet scalar fields $\Phi_1$ and $\Phi_2$, the Lagrangian for the scalar sector of a THDM can 
be written as 
\begin{equation}
\mathcal{L}_{\Phi_{1,2}} = 
(D_{\mu}\Phi_{1})^{\dagger}(D^{\mu}\Phi_{1})+(D_{\mu}\Phi_{2})^{\dagger}(D^{\mu}\Phi_{2})-V(\Phi_{1}, \Phi_{2}),
\end{equation}
where the covariant derivative in standard notation is given as 
$D_\mu = \partial_\mu + i g {\tau^i}/{2}\cdot W_\mu^i + i {g^\prime}/{2} Y B_\mu$, and
the most general renormalisable scalar potential $V(\Phi_{1}, \Phi_{2})$ may be written as:
\begin{eqnarray}
 V(\Phi_1, \Phi_2) = &\,\, m_1^2 \Phi_1^\dag \Phi_1 + m_2^2 \Phi_2^\dag \Phi_2
 - m_3^2 \left(\Phi_1^\dag \Phi_2 + {\rm h.c.} \right) \nonumber \\
 & + \frac{1}{2} \lambda_1 \left(\Phi_1^\dag \Phi_1\right)^2
 + \frac{1}{2} \lambda_2 \left(\Phi_2^\dag \Phi_2\right)^2  
 + \lambda_3 \left(\Phi_1^\dag \Phi_1\right) \left(\Phi_2^\dag \Phi_2\right) \nonumber \\
 & + \lambda_4 \left| \Phi_1^\dag \Phi_2\right |^2 + \frac{1}{2} \lambda_5 \left[  \left( \Phi_1^\dag \Phi_2 \right)^2 
 + {\rm h.c.} \right ] \nonumber \\
 & + \left[ \left[ \lambda_6 \left(\Phi_1^\dag \Phi_1\right) + \lambda_7 \left(\Phi_2^\dag \Phi_2\right)\right]
 \Phi_1^\dag \Phi_2 + {\rm h.c.}\right].
 \label{pot2hdm}
 \end{eqnarray}
 After spontaneous breaking of the electroweak symmetry, five physical Higgs particles are left in the 
 spectrum; one charged Higgs pair, $H^\pm$, one CP-odd scalar, $A$, and two CP-even states,
 $H$ (heaviest) and $h$ (lightest) given as:
 \begin{eqnarray}
 &H^\pm = \sin\beta \,\phi_1^\pm + \cos\beta \,\phi_2^\pm, \\
 &A = \sin\beta \,{\rm Im} \,\phi_1^0 + \cos\beta \,{\rm Im}\, \phi_2^0, \\
 &H = \cos\alpha \left({\rm Re}(\phi_1^0)-v_1\right) + \sin\alpha \left({\rm Re}(\phi_2^0)-v_2\right), \\
 &h = -\sin\alpha \left({\rm Re}(\phi_1^0)-v_1\right) + \cos\alpha \left({\rm Re}(\phi_2^0)-v_2\right).
 \end{eqnarray}
 Here $\phi_i^+$ and $\phi_i^0$ denote the $T_3 = 1/2$ and $T_3 = -1/2$ components of the $i^{th}$ 
 doublet for $i = 1, 2$. The angle $\alpha$ diagonalises the CP-even Higgs squared-mass matrix and
 $\beta$ diagonalises both the CP-odd and charged Higgs sectors with $\tan\beta = v_2/v_1$, where
 $\langle \phi_i^0 \rangle= v_i$ for $i = 1, 2$ and $v_1^2 + v_2^2 \approx (246\,{\rm GeV})^2$. This is 
 a brief summary of how THDMs are constructed. Furthermore, based on different choices of symmetries, 
 couplings to quarks and leptons etc., different models can be built. 
 Models which lead to natural flavour conservation can be named as Type-I, Type-II, Lepton-specific or 
 Flipped 2HDMs, as detailed in Ref.\cite{Branco:2011iw}. 

\subsection{Yukawa interaction model types} 
One of the problems we have in the THDM is that we can't diagonalise the mass matrices simultaneously, 
and this generally cause us to have flavour changing neutral current (FCNCs) at tree level. This 
means that the Yukawa couplings will not be flavour diagonal. The transformation from a general to a 
flavour-conserving THDM can be achieved by imposing a discrete symmetry. This follows from the 
observation that FCNCs vanish at tree level. This results in the THDM having four different model types 
and can be summarised as in Table~\ref{ex}. 
We can write the Yukawa Lagrangian that conserves the FCNCs as follows~\cite{Branco:2011iw}:
\begin{equation*}
\mathcal{L}_{Y}= \bar{Q}_{L,i}(Y_{u,1}^{ij}\tilde{\Phi}_{1}+Y_{u,2}^{ij}\Phi_{2})u_{R,j} +\bar{Q}_{L,i}(Y_{d,1}^{ij}\Phi_{1}+Y_{u,2}^{ij}\tilde{\Phi}_{2})d_{R,j}
\end{equation*}
\begin{equation} 
 \qquad +\bar{L}_{L,i} (Y_{l,1}^{ij}\Phi_{1}+Y_{l,2}^{ij}{\Phi}_{2})l_{R,j} + {\rm h.c.},
\end{equation}
with $\tilde{\Phi}_{i}= -i[\Phi_{i}^{\dagger}\tau_{2}]^{T},  i=1,2$ and $Y_{u,1}^{ij}$ being the $3\times3$ 
Yukawa matrices.
\begin{table}[t]
\caption{\label{ex}Models which lead to natural flavour conservation. The superscript $i$ is a generation index.}
\begin{center}
\begin{tabular}{ccccc}
\br
Models&Type I &Type II &(Leptonic-specific) &(Flipped) \\
\mr
$u_{R}^{i}$&$\Phi_{2}$&$\Phi_{2}$&$\Phi_{2}$&$\Phi_{2}$\\
$d_{R}^{i}$&$\Phi_{2}$&$\Phi_{1}$&$\Phi_{2}$&$\Phi_{1}$\\
$e_{R}^{i}$&$\Phi_{2}$&$\Phi_{1}$&$\Phi_{1}$&$\Phi_{2}$\\
\br
\end{tabular}
\end{center}
\end{table}
After this brief introduction of THDM, we next present the production cross sections of $h$ and $H$ at the
expected energies of future LHeC and FCC-he colliders. 
  
\section{Production of $h$ and $H$ at the LHeC and FCC-he}
\label{collider}
There are various proposals to build new, powerful, high energy $e^+e^-$, $e^-p$ and $pp$ colliders
in the future, where in this proceedings we take an opportunity to probe CP-even scalar Higgs bosons of 
THDM in $e^-\,p$ colliders.
For our studies we consider the LHeC and FCC-he configurations, where depending on electron beam
energies $E_e = 60 - 120$~GeV and the proton beam energy is $E_p = 7$ and $50$~TeV respectively. 
The available centre of mass energies for the LHeC (FCC-he) varies from 
$\sqrt{s} = 2 \sqrt{E_e E_p} \approx 1.3 - 1.8 (3.5 - 5.0)$~TeV.
For more details we refer to Refs.~\cite{AbelleiraFernandez:2012cc, Bruening:2013bga} and more recent
updates can be found on the official website~\cite{lhec}. 
In short these types of colliders
\begin{itemize} 
\item[(i)] provide a clean environment with suppressed backgrounds from the strong interaction processes, and 
free from pile-up, multiple interactions etc.,
\item[(ii)] have asymmetric initial states, where backward and forward scattering can be disentangled and 
\item[(iii)] are known for precision measurements of the dynamical properties of the proton.
\end{itemize}
Studies on Higgs-physics in these configurations can be found in Refs.~\cite{Han:2009pe, Biswal:2012mp, Kumar:2015kca,
Coleppa:2017rgb}. 
It is also important to mention that the THDM has been explored in future $e^+e^-$ machines which can be found
in the International Linear Collider Higgs white paper~\cite{Asner:2013psa} and references there-in.
\begin{figure}
\includegraphics[height=1.6in,width=2.3in]{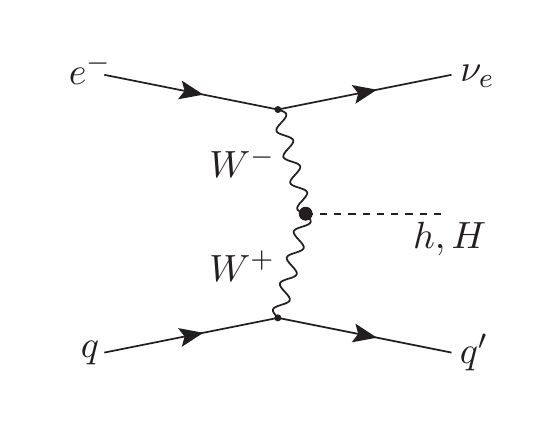}
\includegraphics[height=1.6in,width=2.3in]{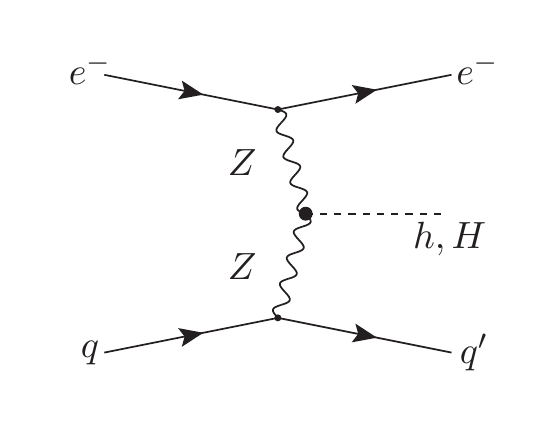}
\caption{Production modes of $h$ and $H$ through charged-current (left) and neutral current (right) 
in $e^- p$ colliders in vector-boson fusion modes. Here $q, q^\prime = u, d, c, s, b, \bar u, \bar d, \bar c, \bar s, \bar b$.}
\label{fig1}
\end{figure}
In $e^-\,p$ configurations, $h$ and $H$ can be produced in two ways as shown in Figure~\ref{fig1}
(here we only show the vector boson fusion diagrams):
\begin{itemize}
\item[(a)] Charged-current (CC): $p \,e^-  \to j \nu_e h/H$ and
\item[(b)] Neutral-current (NC): $p \,e^-  \to j e^-  h/H$,
\end{itemize} 
where $j = u, d, c, s, b, \bar u, \bar d, \bar c, \bar s, \bar b$. For our studies we use the model files for
THDM available in the \texttt{SARAH-4.9.3}~\cite{Staub:2008uz} package, and simulate the events using the 
Monte Carlo event generator package \texttt{MadGraph5-2.4.3.}~\cite{Alwall:2014hca}, where we edit the 
default parameters generated through \texttt{SPheno-3.3.8}~\cite{Porod:2003um, Porod:2011nf} for our 
purpose. Before that we also verified the branchings of $h$ and $H$ to all possible decay modes using the 
\texttt{HDECAY-5.10}~\cite{Djouadi:1997yw} package. For these calculations we used the \texttt{NN23LO1} 
PDF set.

\begin{figure}
\includegraphics[height=2.in,width=3.0in]{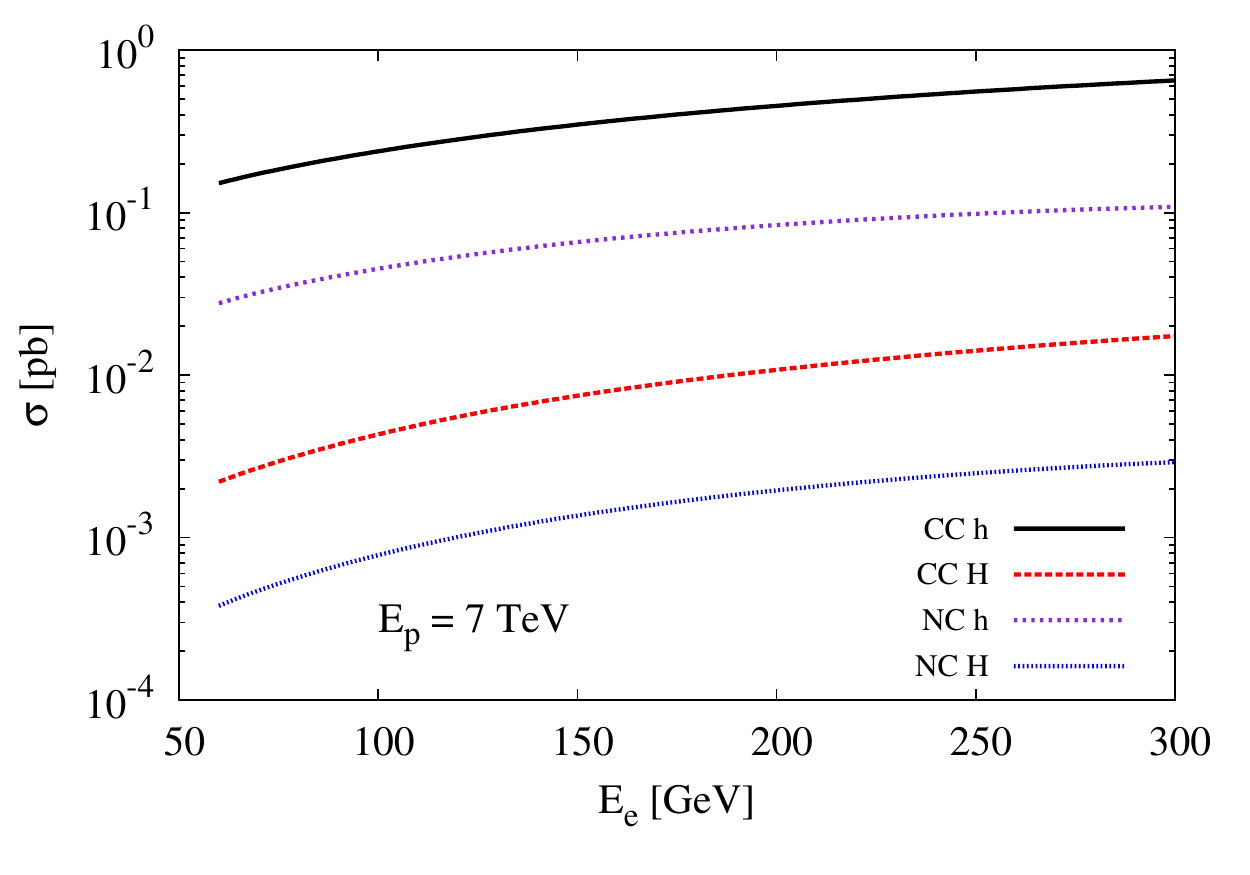}
\includegraphics[height=2.in,width=3.0in]{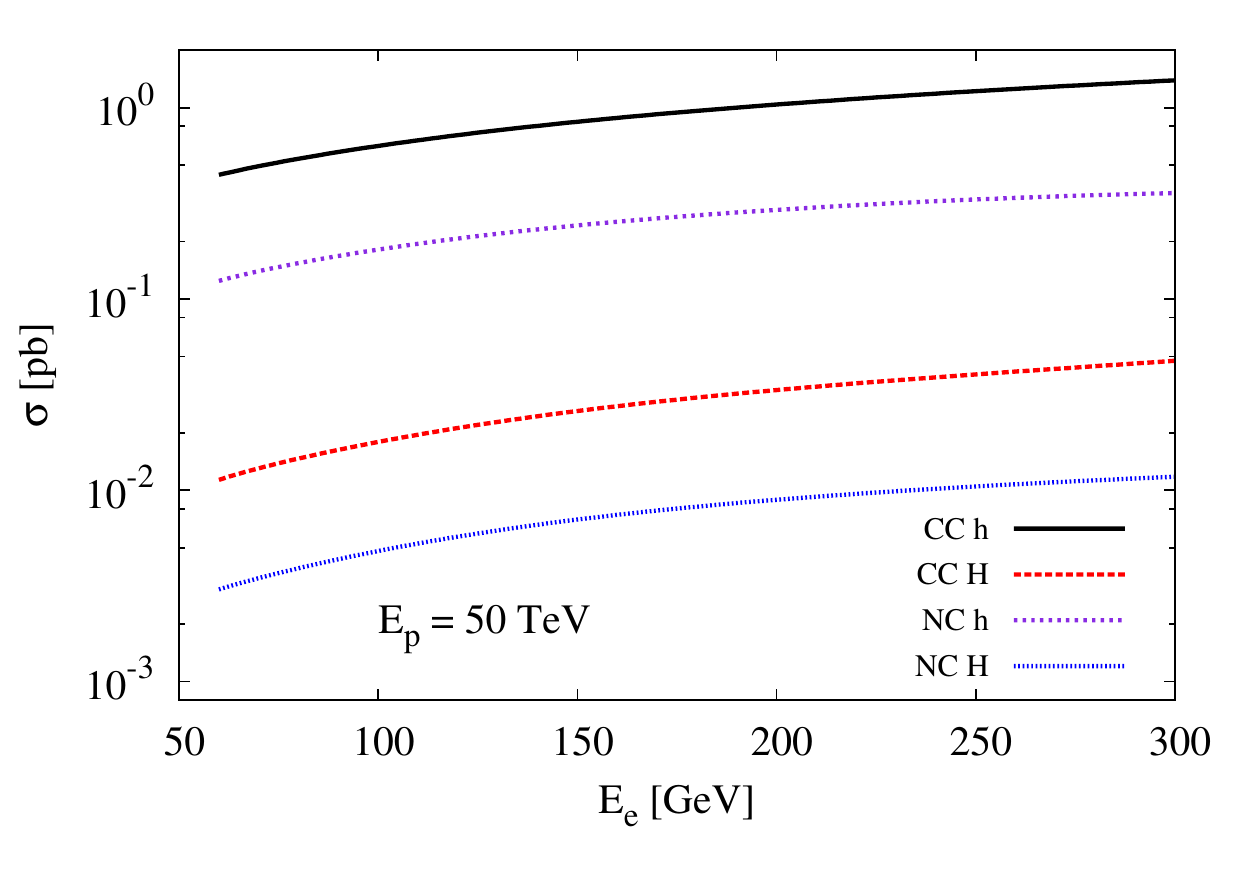}
\caption{Production cross sections of $h$ and $H$ in charged and neutral current THDM Type-I with respect to 
electron beam energies $E_e$ for fixed proton beam energy of $E_p = 7$~TeV (left) and $E_p = 50$~TeV (right).
The default model parameters are as explained in the text with $\tan\beta = 1.0$ and $e^-$ polarisation is taken as
$-80\%$. }
\label{fig2}
\end{figure}

As a preliminary study we consider CP conserving THDM of Type-I, where the default parameters generated 
through \texttt{SPheno} are only real, and mixings are diagonal. In Figure~\ref{fig2} we have shown the total 
production cross sections of $h$ and $H$ with respect to electron beam energies by fixing proton beam 
energies as LHeC and FCC-he recommendations (i.e. 7 and 50 TeV respectively in CC and NC processes 
where $-80\%$ electron polarisation is considered). 
The parameters are:\footnote{As mentioned in the introduction, the choice of $m_H = 270$~GeV is essential here. 
In Refs.~\cite{vonBuddenbrock:2015ema,vonBuddenbrock:2016rmr,Kumar:2016vut}
the authors investigated a number of final states with  the 7 and 8~TeV data at ATLAS and CMS. This included 
the Higgs boson transverse momentum, the limits on the production of di-Higgs bosons, the invariant mass of 
$VV (V=Z,W^\pm)$ and the results of the search for associated top-Higgs ($tth$) production. Features in the 
data could be explained by a hypothetical scalar-boson of mass around 270~GeV.}
\begin{itemize}
\item[(a)] $m_h = 125$~GeV, $m_H = 270$~GeV, $m_A = 450$~GeV, $m_{H^\pm} = 400$~GeV 
(as recommended in Refs.~\cite{vonBuddenbrock:2015ema, vonBuddenbrock:2016rmr, Kumar:2016vut}),
\item[(b)] $\tan\beta = 1.0$, $\alpha = -0.53$ and
\item[(c)] $\lambda_1 = 0.1$, $\lambda_2 = 0.27$, $\lambda_3 = 1.1$, $\lambda_4 = - 0.5$, $\lambda_5 = 0.5$.
\end{itemize}
Here it is important to mention that the $hVV$ ($V = W^\pm, Z$) couplings in the THDM is dependent on the choice
of the mixing angles $\alpha$ and $\beta$. For example, couplings for $hW^+W^-, hZZ$ depend on $\sin(\beta-\alpha)$, 
while in the case of $HW^+W^-, HZZ$, it is proportional to $\cos(\beta-\alpha)$ etc. Note that in the SM there is no
such mixing angle dependence on such couplings. In our choice of parameters $\cos(\beta-\alpha) = \pm 0.25$ and
$\sin(\beta-\alpha) = 0.97$, which can further be modified and scanned according to the constraints from the ATLAS and
CMS analyses at the LHC. Here we chose these values so as to not evade the cross section limits as given in 13 TeV
data from ATLAS~\cite{ATLAS:2016oum}, where a search for new heavy scalar bosons into the four-lepton final state 
is performed by exploring the mass range [200-1000]~GeV. 
In Table~\ref{branch} we estimate the branching ratio (BR) of $h, H, A$ and $H^\pm$ using the \texttt{HDECAY-5.10} package 
with these parameter choices. The dominant BRs are shown in bold and suggest the possibility for appropriate 
final states to look for signal to background estimations. 
 
Interestingly one of the important expectations from the LHeC/FCC-he is that we should be able to measure couplings 
of $H$ with a very good signal to background ratio when compared with $pp$ collisions. In our ongoing work we estimate 
the appropriate signal to background ratio of $H$ production in CC and NC processes, so that we can have an idea 
for the level of the BR that can be accessible in these types of machines (which are highly dependent on the THDM parameters).
For instance, from Figure~\ref{fig1} for $E_e=60$~GeV the CC (NC), $H$ production is about 2.2 fb (0.38 fb), which will 
give us about 2000 (380) $H$ produced for 1 ab$^{-1}$ integrated luminosity at the LHeC. 
Here it is important to mention that in Ref.~\cite{Han:2009pe} the SM Higgs-boson BR to the $b\bar b$ mode has 
been studied, and tagging the forward jet assures this decay mode has a significantly reduced background which
improves the purity of the signal. Therefore, estimations of the BRs (shown in Table~\ref{branch}) suggest 
$H \to h h \to b\bar b b\bar b$ as the dominant mode (which would be a signal rate of $\sim 55$\% of the total production cross 
section of $H$, where the rate of one of the dominant background $b\bar b b\bar b j$ in the CC (NC) at the LHeC would be
$\sim 2.0$~fb ($22$~fb) while for the FCC-he the cross section is around 3 (7.5) times higher for $E_e = 60$~GeV), 
while $H \to h h \to W^+ W^-$ are subdominant signal modes.
Hence the sensitivity of the $Hhh, HW^+W^-$ and $HZZ$ couplings in this mode can be measured at the LHeC/FCC-he, 
as discussed in Ref.~\cite{Kumar:2015kca}.  
\begin{table}[t]
\caption{\label{branch}Branching ratio of $h, H, A$ and $H^\pm$ by considering the parameter choices as: 
 $m_h = 125$~GeV, $m_H = 270$~GeV, $m_A = 450$~GeV and $m_{H^\pm} = 400$~GeV, $\tan\beta = 1$,
 $\alpha = -0.53$, $\lambda_1 = 0.1$, $\lambda_2 = 0.27$, $\lambda_3 = 1.1$,
 $\lambda_4 = -0.5$ and $\lambda_5 = 0.5$ for THDM Type-I. (Dominant BRs are shown in bold)}
\begin{center}
\begin{tabular}{lc|lc|lc|cc}
\br
Modes & $h$ & Modes & $H$ & Modes	& $A$ & Modes & $H^\pm$ \\
\mr
$\bf{b\bar b}$	        	& $\bf{6.5 \times 10^{-1}}$ &$b\bar b$	        		& $6.8 \times 10^{-4}$ 	&$b\bar b$	& $2.7 \times 10^{-4}$ 	 & $bc$        & $5.9 \times 10^{-7}$ 	\\
$\bf{\tau^+\tau^-}$	& $\bf{7.0 \times 10^{-2}}$ &$\tau^+\tau^-$		& $8.5 \times 10^{-5}$ 	&$\tau^+\tau^-$	& $3.8 \times 10^{-5}$ 	 & $\tau\nu$ & $4.6 \times 10^{-5}$ 	\\
$\mu^+\mu^-$		& $2.5 \times 10^{-4}$ 	&$\mu^+\mu^-$		& $3.0 \times 10^{-7}$ 	&$\mu^+\mu^-$& $1.3 \times 10^{-7}$ 	 & $\mu\nu$ & $1.6 \times 10^{-7}$ 	\\
$s\bar s$	                 & $2.5 \times 10^{-4}$ 	&$s\bar s$	                 & $2.6 \times 10^{-7}$ 	&$s\bar s$	& $9.6 \times 10^{-8}$ 	 & $su$	& $6.2 \times 10^{-9}$ 	\\
$\bf{c\bar c}$	        & $\bf{3.2 \times 10^{-2}}$ &$c\bar c$	                 & $3.3 \times 10^{-5}$ 	&$c\bar c$	& $1.4 \times 10^{-5}$ 	 & $cs$	& $1.5 \times 10^{-5}$ 	\\
$t\bar t$	                 & $0.0 \times 10^{-0}$ 	&$t\bar t$	                 & $8.5 \times 10^{-7}$ 	&$\bf{t\bar t}$	& $\bf{7.6 \times 10^{-1}}$ & $\bf{tb}$& $\bf{8.7 \times 10^{-1}}$ 	\\
$\bf{gg}$          		& $\bf{8.5 \times 10^{-2}}$ &$gg$          		& $5.5 \times 10^{-4}$ 	&$gg$		& $3.1 \times 10^{-3}$ 	 & $cd$	& $8.2 \times 10^{-7}$ 	\\
$\gamma\gamma$	& $1.4 \times 10^{-3}$ 	&$\gamma\gamma$	& $6.7 \times 10^{-6}$ 	&$\gamma\gamma$& $9.4 \times 10^{-6}$ & $bu$	& $4.1 \times 10^{-9}$ 	\\
$Z\gamma$		& $1.0 \times 10^{-3}$ 	&$Z\gamma$		& $1.1 \times 10^{-5}$ 	&$Z\gamma$	& $2.4 \times 10^{-6}$	  & $ts$	& $1.4 \times 10^{-3}$ 	\\
$\bf{W^+W^-}$		& $\bf{1.4 \times 10^{-1}}$ &$\bf{W^+W^-}$		& $\bf{7.1 \times 10^{-2}}$ &$\bf{Zh}$	& $\bf{5.1 \times 10^{-2}}$   & $td$	& $6.5 \times 10^{-5}$ 	\\
$\bf{ZZ}$			& $\bf{1.8 \times 10^{-2}}$ &$\bf{ZZ}$			& $\bf{3.1 \times 10^{-2}}$ &$\bf{ZH}$	& $\bf{1.84 \times 10^{-1}}$ & $\bf{hW^\pm}$& $\bf{4.6 \times 10^{-2}}$ 	\\
				&  					&$\bf{hh}$			& $\bf{9.0 \times 10^{-1}}$ &$W^+H^-$	& $3.6 \times 10^{-5}$ 	   & $\bf{HW^\pm}$& $\bf{8.3 \times 10^{-2}}$  	\\
\br
\end{tabular}
\end{center}
\end{table}

\section{Summary and future work}
\label{future}
In this proceedings we have provided a preliminary study of the cross section measurements of $h$ and $H$ using 
energies from the LHeC and FCC-he configurations. For $E_e = 60$~GeV and $E_p = 7 (50)$~TeV the chosen
parameters give substantial signal rates for $H$ production in CC and NC modes (with $m_H = 270$), 
$\sigma_H^{\rm CC} \sim 2.2 (11.3)$~fb and $\sigma_H^{\rm NC} \sim 0.38 (3)$~fb respectively. 
While for $h$ production these rates are $\sigma_h^{\rm CC} \sim 150 (450)$~fb and 
$\sigma_h^{\rm NC} \sim 28 (124)$~fb.
Here we considered $e^-$ polarisation of $-80\%$ which enhances the un-polarised signal cross sections (and
also backgrounds) by 1.8 times. We also gave estimations of the BRs for various decay modes
of $h, H, A$ and $H^\pm$, based on our parameter choice, and discussed its implications.
For these parameter choice we discussed dominant final state such as, $H \to h h \to b\bar b b \bar b$
and thus at the LHeC/FCC-he one can estimate the sensitivity of $Hhh$ couplings including the $HW^+W^-$
and $HZZ$.
  
There are various analyses which remain for our future studies, some of these are:
\begin{itemize}
\item[(1)] A full analysis of the kinematics of the final states using decay modes of $h$ and $H$ with appropriate
backgrounds and an estimation of the accuracy of the BRs in different final states, say, $b\bar b, \tau^+ \tau^-$ etc.
\item[(2)] A study of the dependence of the parameter space on different observables. It is important to mention
here that the choice of parameter space must be consistent with all current Higgs constraints by using
\texttt{HiggsSignal} and \texttt{HiggsBounds}~\cite{higgs-sig-bound}. Though the single parameter choice we took for our
studies are also consistent with these constraints.
\item[(3)] An exploration of the other scalars of the THDMs, such as the CP-odd $A$ and the charged $H^\pm$
should be added for future studies. Note that few studies for light $H^\pm$ production and decay in $e^- p$ environment 
can be found in Refs.~\cite{Moretti:1997ip, Hernandez-Sanchez:2016vys}.
\item[(4)] A consistency check with the constrained parameter space of the THDMs through various
studies available at the LHC and $e^+e^-$ colliders.
\item[(5)] Studies related to the FCNCs, which are also very important.
\end{itemize}

\end{document}